\documentclass{an}
\usepackage{times}
\usepackage{graphicx}
\usepackage{fancyhdr}
\usepackage{url}
\usepackage{bm}
\graphicspath{{./fig/}{./png/}}
\newcommand{\EQ}{\begin{equation}}
\newcommand{\EN}{\end{equation}}
\newcommand{\EQA}{\begin{eqnarray}}
\newcommand{\ENA}{\end{eqnarray}}
\newcommand{\eq}[1]{(\ref{#1})}

\newcommand{\Eq}[1]{Eq.~(\ref{#1})}

\newcommand{\Fig}[1]{Fig.~\ref{#1}}
\newcommand{\FFig}[1]{Figure~\ref{#1}}

\newcommand{\bra}[1]{\langle #1\rangle}

\newcommand{\meanAA}{\overline{\vec{A}}}
\newcommand{\meanBB}{\overline{\vec{B}}}
\newcommand{\meanJJ}{\overline{\vec{J}}}

\newcommand{\BB}{{\vec{B}}}
\newcommand{\JJ}{{\vec{J}}}
\newcommand{\jj}{{\vec{j}}}
\newcommand{\AAA}{{\vec{A}}}
\newcommand{\aaaa}{{\vec{a}}}
\newcommand{\bb}{{\vec{b}}}

\newcommand{\nab}{\mbox{\boldmath $\nabla$} {}}

\newcommand{\dd}{{\rm d} {}}

\newcommand{\Mm}{\,{\rm Mm}}

\newcommand{\yapj}[3]{: #1, {ApJ} {#2}, #3}

\newcommand{\yan}[3]{: #1, {AN} {#2}, #3}

\newcommand{\yana}[3]{: #1, {A\&A} {#2}, #3}

\newcommand{\ypre}[3]{: #1, {PhRvE} {#2}, #3}

\newcommand{\yjour}[4]{: #1, {#2} {#3}, #4}

\newcommand{\ybook}[3]{: #1, {\it #2} (#3)}

\begin{document}

\title{The helicity constraint in spherical shell dynamos}
\authorrunning{A. Brandenburg et al.}
\author{A. Brandenburg\inst{1}, P. J. K\"apyl\"a\inst{1,2},
D. Mitra\inst{3}, D. Moss\inst{4}, R. Tavakol\inst{3}
}
\institute{
NORDITA, Roslagstullsbacken 23, SE-10691 Stockholm, Sweden
\and
Observatory, University of Helsinki, PO Box 14, FI-00014 University of Helsinki, Finland
\and
School of Mathematical Sciences, Queen Mary, University of London, Mile End Road, London E1 4NS, UK
\and
School of Mathematics, University of Manchester, Oxford Road, Manchester M13 9PL, U.K.
}

\date{\today}

\abstract{
The motivation for considering distributed large scale dynamos
in the solar context
is reviewed in connection with the magnetic helicity constraint.
Preliminary accounts of 3-dimensional direct numerical simulations (in spherical shell segments)
and simulations of 2-dimensional mean field models (in spherical shells) 
are presented. Interesting similarities as well as some 
differences are noted.
}

\maketitle

\section{The context of the solar dynamo}

At the moment we do not really know the location of the solar dynamo.
It is widely assumed that it is located at the bottom of the convection
zone, or that at least its toroidal field resides mostly at the bottom.
However, this may not be the case, and instead a major part of the
toroidal field may reside in the bulk of the convection zone -- possibly even
as high up as the near-surface shear layer, corresponding to the outer
$35\Mm$ of the Sun (Brandenburg 2005).
In this introductory section we discuss certain problems and properties
associated with this proposal.

A particularly exciting possibility is to associate the equatorward migration
of sunspot activity with the negative radial shear in the near-surface shear layer.
This, together with a positive $\alpha$ effect in the northern hemisphere,
could, according to standard mean-field theory (Krause \& R\"adler 1980),
give rise to equatorward migration of the mean field.
An obvious problem associated with this proposal is the fact that the
near-surface shear layer is rather thin, and so the resulting aspect ratio
of the container would favour solutions with many toroidal flux belts
in one hemisphere (Moss et al.\ 1990).

Another potential problem might be the fact that the local turnover time
in the near-surface shear layer is rather short (1 day compared to 12 days
in the lower part of the convection zone).
This is sometimes thought to be a difficulty if one wants to explain cycle
times many times longer than this.
However, it may not be justified to draw such a conclusion based
on conventional mean field theory that ignores the effects of magnetic
helicity conservation.
This is true not only for near-surface shear layer dynamos, but even
for what is now usually referred to as the Babcock-Leighton dynamo
(Dikpati \& Charbonneau 1999),
which is based on a nonlocal $\alpha$ effect (the field at the bottom
of the convection zone affects the electromagnetic force at the surface);
see Brandenburg \& K\"apyl\"a (2007) for an explicit demonstration.
Even mean-field effects other than those based on an $\alpha$ effect,
e.g.\ the so-called shear--current effect of Rogachevskii \& Kleeorin
(2003), produce magnetic helicity in the mean field and tend to generate
also magnetic helicity of suitable sign in the small scale field
in order to
quench the dynamo.
The latter drives a magnetic $\alpha$ effect that affects the mean
electromotive force in a way which quenches the shear--current effect
even if there is no kinetic $\alpha$ effect in the usual sense
(Brandenburg \& Subramanian 2005).

Quite generally, when the magnetic Reynolds number is large, mean field
dynamos are only able to operate on time scales faster than the resistive
time provided the resulting small scale magnetic helicity is shed from
the dynamo domain.
That the Sun sheds significant amounts of magnetic helicity through
active regions and coronal mass ejections is now well known
(D\'emoulin et al.\ 2002).
Thus, it seems that an important ingredient of any solar dynamo model
should be a penetrable surface that allows magnetic helicity to escape
from the dynamo domain.
An important aim of future work is therefore to improve our understanding
of dynamos that are controlled by magnetic helicity fluxes.

Considerable progress has recently been made by using direct simulations
of the solar dynamo.
The simulations of Brun et al.\ (2004) and Browning et al.\ (2006)
suggest that dynamo action in the
lower overshoot layer leads to cyclic reversals.
However, there is at present no clear evidence for migratory behaviour.
On the other hand, current mean field modelling assumes that the magnetic
field propagates equatorward due to the meridional circulation that
flows equatorward at the bottom of the convection zone
(e.g.\ Dikpati \& Charbonneau 1999).
However, this approach only works by assuming that the {\it turbulent}
magnetic Prandtl number is about 100, which is hard to understand,
because both theory and simulations suggest that this value should
be around unity (Yousef et al.\ 2003).
More importantly, circulation--dominated models neglect the effects
of small scale magnetic helicity that contributes to the quenching
of dynamo action.
It is therefore useful to establish details of magnetic helicity
conservation in spherical geometry.
In the present work we consider both direct three-dimensional simulations
and two-dimensional mean field models.

\section{The magnetic helicity constraint}

In order to make contact between what has been learnt from Cartesian
and spherical systems it is important to start with simple cases.
This means closed (perfectly conducting) boundaries and fully helical
turbulence in homogeneous (or nearly homogeneous) domains.
At the level of mean field theory, this means a 
spatially constant $\alpha$ effect.

Of course, real systems are not uniform and the $\alpha$ effect changes
sign across the equator.
Nevertheless, for testing and illustrative purposes it is useful to
consider a uniform $\alpha$ effect in spherical geometry
(Krause \& R\"adler 1980).
If there is only the $\alpha$ effect and turbulent diffusivity,
$\eta_{\rm t}$, in addition to microscopic diffusivity, $\eta$,
the mean magnetic field, $\meanBB$, is governed by the equation
\EQ
{\partial\meanBB\over\partial t}
=\nab\times\alpha\meanBB+\eta_{\rm T}\nabla^2\meanBB,
\label{mfmodel}
\EN
where $\eta_{\rm T}=\eta_{\rm t}+\eta$ is the total magnetic diffusivity.
(Both $\alpha$ and $\eta_{\rm T}$ are in this section assumed 
to be constants.)
In a Cartesian domain of size $L$ the critical value of $\alpha$
for dynamo action (i.e.\ exponentially growing solutions) is
\EQ
\alpha/\eta_{\rm T}>k_1\equiv 2\pi/L.
\EN
For a full sphere the critical value is related to the first zero
of the lowest order spherical Bessel function, i.e.\
\EQ
\alpha/\eta_{\rm T}>k_{\rm1,eff}\equiv 4.49/R.
\label{k1eff}
\EN
In the present paper
we are particularly interested in perfectly conducting boundary
conditions, because in that case the magnetic helicity is conserved.
This fact yields an interesting quantitative connection between the
magnetic energies contained in large scale and small scale fields,
as will be explained in the following.
In a closed domain whose boundaries are not crossed by any field lines the
evolution equation of magnetic helicity is given by
\EQ
{\dd\over\dd t}\int \AAA\cdot\BB\;\dd V=
-2\eta\int \JJ\cdot\BB\;\dd V,
\label{dABdt}
\EN
where $\JJ=\nab\times\BB$ is the current density (in units where the
permeability is unity) and $\BB=\nab\times\AAA$ is the magnetic field
expressed as the curl of the magnetic vector potential.
The integral on the rhs of \Eq{dABdt} is referred to as the current helicity.
Remarkably, in the steady state we then have
\EQ
\int \JJ\cdot\BB\;\dd V=0,
\label{curhel}
\EN
even though there can be strong driving of current helicity due to the helical
nature of the forcing.

Such forcing produces primarily helical fields at the scale of the
driving (we call the associated wavenumber $k_{\rm f}$).
If this scale is small compared with the system size (associated
wavenumber $k_1$), we can have current helicity of finite
magnitude at small and large scales.

To elaborate on this further, we define mean fields by averaging over
one or two coordinate directions and denote the result by an overbar,
i.e.\ we have $\meanJJ=\nab\times\meanBB$ and $\meanBB=\nab\times\meanAA$.
The corresponding fluctuating fields are indicated by lower case symbols,
i.e.\ $\bb=\BB-\meanBB$, etc., so the idea is then that
\EQ
\bra{\meanJJ\cdot\meanBB}=-\bra{\jj\cdot\bb}\neq0,
\label{JBjb}
\EN
where angular brackets indicate volume averages.
The wave numbers (inverse length scales) associated with large and
small scale fields, $k_{\rm m}$ and $k_{\rm f}$, respectively,
can be given as
\EQ
k_{\rm m}^2\equiv\bra{\meanJJ\cdot\meanBB}/\bra{\meanAA\cdot\meanBB},\quad
k_{\rm f}^2\equiv\bra{\jj\cdot\bb}/\bra{\aaaa\cdot\bb}.
\EN
For a fully helical field we also have
$\bra{\meanJJ\cdot\meanBB}=\pm k_{\rm m}\bra{\meanBB^2}$,
$\bra{\jj\cdot\bb}=\mp k_{\rm f}\bra{\bb^2}$,
where the signs depend on the sign of the helicity.
If the large and small small scale fields are not fully helical
we have instead
\EQ
\epsilon_{\rm m}\equiv\bra{\meanJJ\cdot\meanBB}/(k_{\rm m}\bra{\meanBB^2}),\quad
\epsilon_{\rm f}\equiv-\bra{\jj\cdot\bb}/(k_{\rm f}\bra{\bb^2}),
\EN
where $\epsilon_{\rm m}$ and $\epsilon_{\rm f}$ quantify the fractions
of the large and small scale fields that are helical.
Inserting this into \Eq{JBjb} yields
\EQ
\bra{\meanBB^2}
={\epsilon_{\rm f}k_{\rm f}\over\epsilon_{\rm m}k_{\rm m}}\bra{\bb^2},
\EN
which shows that the saturation amplitude of the mean field is proportional
to the ratio of the scale separation, i.e.\ $k_{\rm f}/k_{\rm m}$, and that
lowering the
fractional helicity of the small scale field lowers the saturation
amplitude of the large scale field.
Moreover, lowering the fractional helicity of the large scale field 
actually enhances the saturation field strength.

All these aspects have been confirmed in Cartesian domains using numerical
simulations; see Brandenburg \& Subramanian (2005) for a review.
Most surprising is perhaps the last aspect, i.e.\ that lowering the
fractional helicity
of the large scale field ($|\epsilon_{\rm m}|<1$) increases the saturation
field strength.
Simulations in a Cartesian box, where the boundaries in one direction
are not periodic but perfectly conducting, show that the energy of the
large scale magnetic field can exceed the kinetic energy (which is approximately
the energy of the small scale field, denoted by $B_{\rm eq}^2$)
by a factor that is equal to the
scale separation ratio $k_{\rm f}/k_{\rm m}$ (Brandenburg 2001).

In the following we give a brief preliminary discussion 
of both helically forced simulations in spherical shell segments
as well as mean field models in spherical shells.

\begin{figure*}[t!]\begin{center}
\includegraphics[width=\textwidth]{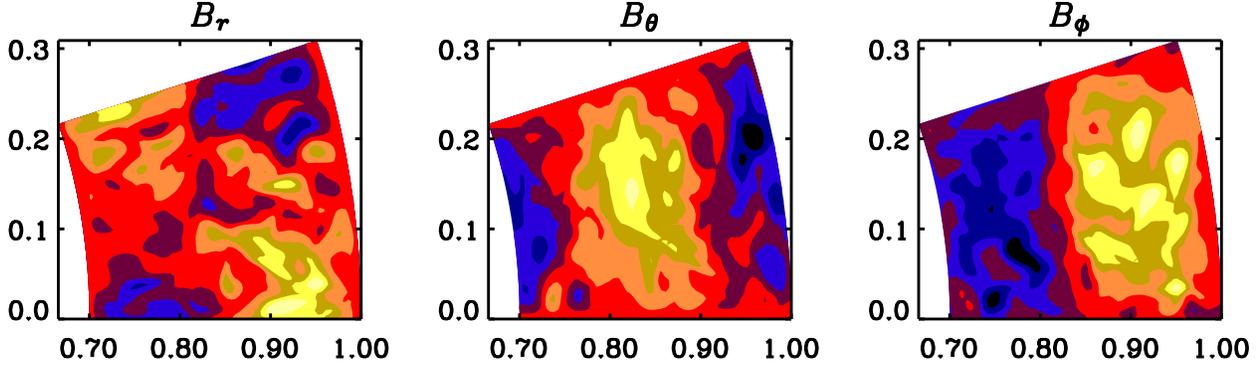}
\end{center}\caption[]{ 
Visualization of the three magnetic field components in an arbitrarily
chosen meridional plane for a turbulent
dynamo in a spherical segment with fully helical forcing
with $R_{\rm m}=30$ (see text for boundary conditions).
}\label{pbb3}\end{figure*}

\section{Turbulence in spherical shell segments}

Three dimensional turbulence simulations in spherical shell segments have been performed
using an experimental version of the \textsc{Pencil Code}\footnote{
\url{http://www.nordita.org/software/pencil-code}.}, which is a
non-conservative, high-order, finite-difference code (sixth order in
space and third order in time) for solving the compressible hydrodynamic
equations.

In \Fig{pbb3} we show a visualization of the three components of 
a snapshot of the magnetic field after it has reached
saturation. In this run the magnetic Reynolds number is $R_{\rm m}=30$,
where
\EQ
R_{\rm m}=u_{\rm rms}/(\eta k_{\rm f}),
\EN
with $k_{\rm f}/k_{\rm m}=5$, and $k_{\rm m}=k_1$ with
\EQ
k_1=2\pi/(R-R_{\rm in})
\label{k1}
\EN
where $R-R_{\rm in}$ is the the thickness of the shell whose
inner and outer radii are given by $R$ and $R_{\rm in}$.
In the simulations presented here we have used $R_{\rm in}=0.7\,R$, 
which gives $k_1=21$ and $k_{\rm f}=105$.
The latitudinal and longitudinal extent of the domain is $\pi/10$,
so the domain is roughly cubical.
This run uses periodic boundary condition along the azimuthal direction
and perfectly conducting boundary conditions along the other two.

The azimuthal component, $B_\phi$, is particularly prominent and shows a
segregation of large scale positive and negative components in the radial direction,
and a pattern in the latitudinal component that is shifted in the radial
direction by 1/4 of the thickness of the shell.

\begin{figure}[t!]\begin{center}
\includegraphics[width=\columnwidth]{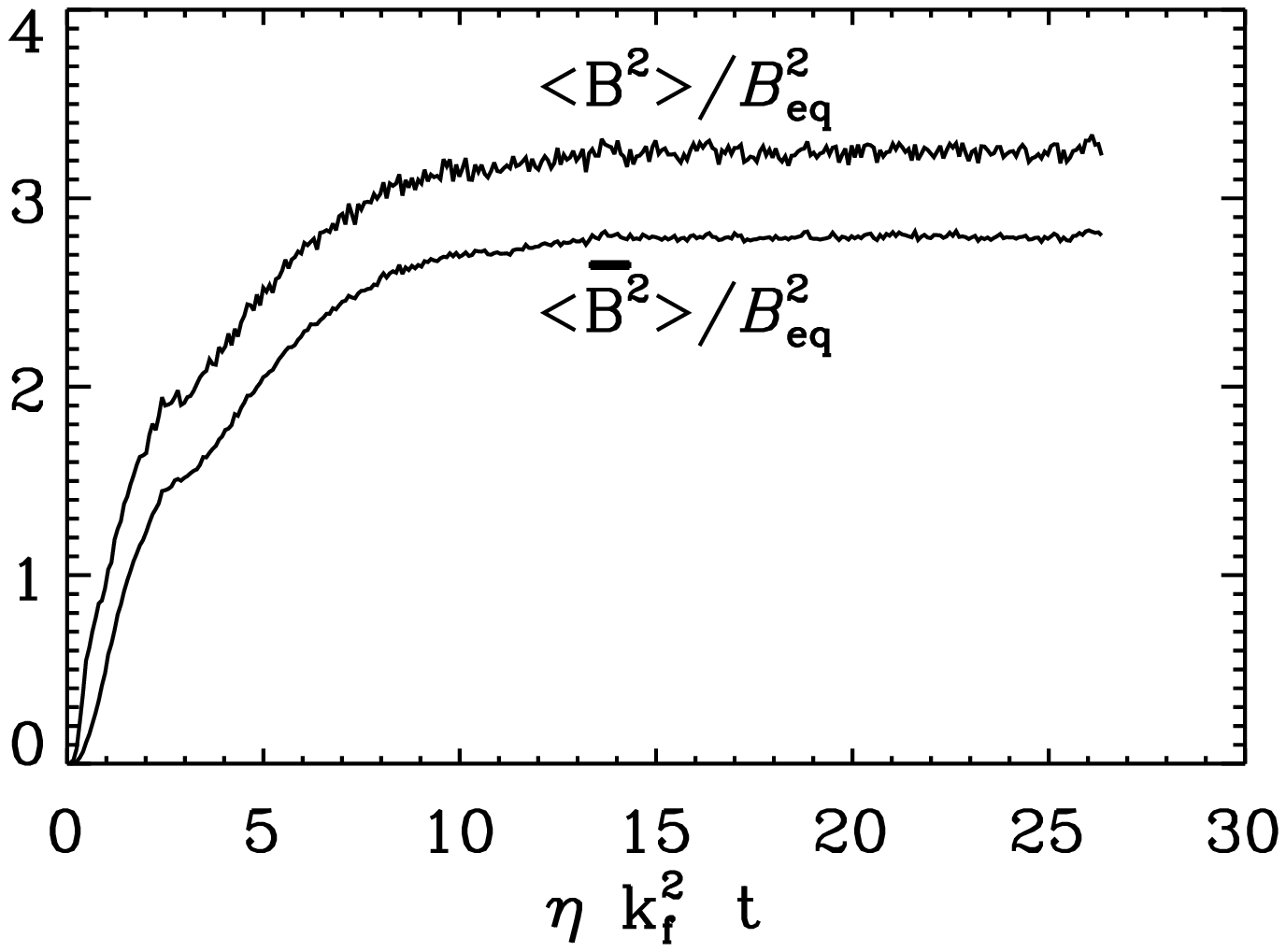}
\end{center}\caption[]{ 
Saturation behaviour of the normalized magnetic energies of the total
field, $\bra{\BB^2}$, and the mean field, $\bra{\meanBB^2}$, in a
three-dimensional turbulence simulation
with $R_{\rm m}=30$, i.e.\ the same run as in \Fig{pbb3}.
}\label{pmeanfield_t}\end{figure}

The temporal evolution of the magnetic field is shown in \Fig{pmeanfield_t}.
The energy of the total magnetic field exceeds the kinetic energy
by about a factor of $3$, most of which is contained in the mean field. 
These results are similar to behaviours observed in 
simulations in periodic Cartesian domains (see Brandenburg 2001).

\section{Mean field models}

We next consider 2-dimensional axisymmetric simulations of \Eq{mfmodel} in spherical shells, but
with a dynamical $\alpha$ effect that obeys its own time-dependent
differential equation,
\EQ
{\partial\alpha\over\partial t}=
-2\eta_{\rm t}k_{\rm f}^2
\left({\alpha\meanBB^2-\eta_{\rm t}^2\meanJJ\cdot\meanBB\over B_{\rm eq}^2}
+{\alpha-\alpha_{\rm K}\over R_{\rm m}}\right).
\EN
(Kleeorin et al.\ 1995, Blackman \& Brandenburg (2002).
The mean field models were computed using a modified version of the 
finite-difference code described in K\"apyl\"a et al. (2006).
Note that $\alpha$ is now no longer constant, but it is quenched locally
relative to the kinematic value, $\alpha_{\rm K}$, which is still
spatially constant.
In closed domains, such as those considered in this paper,
the amount of quenching depends sensitively on the value of
the magnetic Reynolds number, $R_{\rm m}\equiv\eta_{\rm t}/\eta$.
Note also that there are no additional free parameters emerging from
this type of nonlinearity.

\begin{figure}[t!]\begin{center}
\includegraphics[width=\columnwidth]{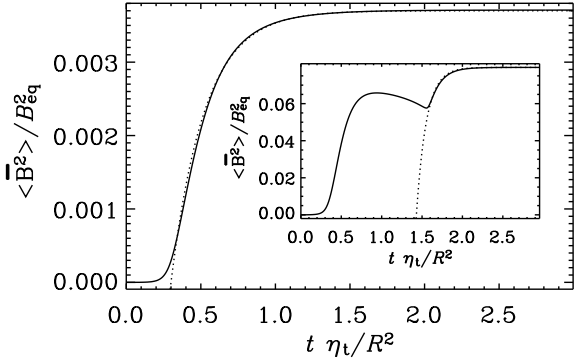}
\end{center}\caption[]{ 
Saturation behaviour of a mean field model with perfect conductor boundary
condition for $R_{\rm m}=1000$, $k_{\rm f}R=100$, and $R_{\rm in}=0.5\,R$.
The inset shows the result for $R_{\rm m}=30$ and $R_{\rm in}=0.7\,R$.
Dashed lines give the fit \eq{fit}.
}\label{pb2fit}\end{figure}

We begin by considering first the saturation behaviour of the
magnetic field in such a model.
\FFig{pb2fit} shows the evolution of $\bra{\meanBB^2}/B_{\rm eq}^2$ for
a model with $R_{\rm m}=1000$, $k_{\rm f}R=100$, and $R_{\rm in}=0.5\,R$,
as well as a case with $R_{\rm m}=30$ and $R_{\rm in}=0.7\,R$ (inset).
Also shown is the fit
\EQ
{\bra{\meanBB^2}\over B_{\rm eq}^2}
={\epsilon_{\rm f}k_{\rm f}\over\epsilon_{\rm m}k_{\rm m}}
\left[1-e^{-2\eta k_{\rm1,eff}^2(t-t_{\rm sat})}\right],
\label{fit}
\EN
where $t_{\rm sat}$ is approximately the time when the small scale field
has saturated, and $k_{\rm1,eff}$ is a new effective wavenumber
that might be related to that defined in \Eq{k1eff}, or even $k_1$
defined in \Eq{k1}, but otherwise this is just treated as a fit parameter.
For details regarding the derivation of this fit formula
see Brandenburg (2001).

\begin{figure}[t!]\begin{center}
\includegraphics[width=\columnwidth]{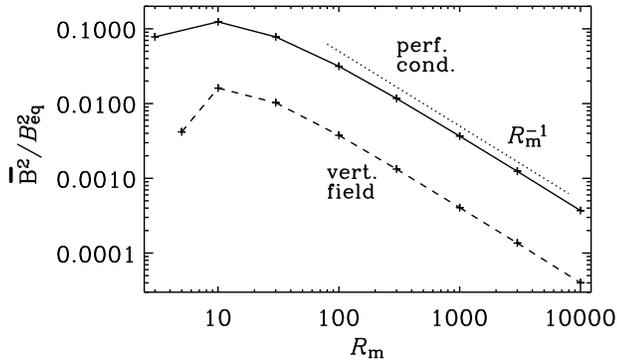}
\end{center}\caption[]{ 
Comparison of the saturation behaviours of a mean field model
for perfect conductor (solid line) and vertical field (broken)
boundary conditions.
}\label{pspherical_model_pc}\end{figure}

It turns out that, as the value of $R_{\rm m}$ is increased, the final
saturation field strength of the large scale field decreases;
see \Fig{pspherical_model_pc}.
This behaviour is familiar in the case of open boundary conditions,
but it is unexpected in the case of perfectly conducting boundaries.
The only difference between these two cases is that the energy is larger 
by 
a factor of about 10 when the boundaries are perfectly conducting.

\begin{figure}[t!]\begin{center}
\includegraphics[width=\columnwidth]{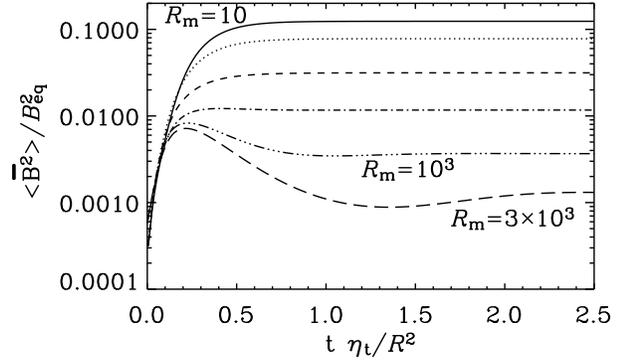}
\end{center}\caption[]{ 
Saturation behaviour of a mean field model with perfect conductor boundary
condition with different values of $R_{\rm m}$,
$k_{\rm f}R=300$, and $R_{\rm in}=0.5\,R$.
}\label{pb2ts_pc}\end{figure}

Also depicted in \Fig{pb2ts_pc} are the time traces of $\bra{\meanBB^2}/B_{\rm eq}^2$ 
for different values of $R_{\rm m}$.
Note that for models with $R_{\rm m}>300$ there is a tendency for a
decline of  $\bra{\meanBB^2}$ after having reached an initial maximum.

\section{Conclusions}

We have briefly presented preliminary results comparing direct simulations 
in spherical shell segments with their Cartesian analogues,
on the one hand, and mean field models in spherical shells,
on the other.

Simulations in Cartesian boxes and spherical
shell segments both show qualitatively similar features
in the mean magnetic field as well as similar saturation
behaviour in the normalized mean magnetic field.
On the other hand the saturation value is lower
in mean field models than in the three-dimensional simulations.

Clearly more work is needed to clarify further  
the relation between these models,
before we can proceed to the arguably more realistic case with a penetrable
boundary. Work is in progress
in this direction and will be reported elsewhere.

\acknowledgements
{PJK acknowledges support from the
Helsingin Sanomat foundation. DM is
supported by Leverhulme trust.
Computational resources were granted 
by CSC (Espoo, Finland) and QMUL HPC facilities purchased
under the SRIF initiative.}

\end{document}